\input harvmac
\input epsf

\def\ap{\alpha'}
\def\half{{1\over 2}}

\def\sh{\hat{\sigma}}

\def\({\left(}
\def\){\right)}
%%%%%%%%%%%%%%%%%%%%%%%%%%%%%%%%%%%%%%%%%%%%%%%%%%%%%%%%%%%%%%%%%%%%%

\Title{}{\vbox{\centerline{Brane Inflation After WMAP Three Year
Results}}}

\centerline{Qing-Guo Huang $^{1,2}$, Miao Li$^{1,2}$ and
Jian-Huang She$^{1,3}$}
\bigskip \centerline{\it $^1$Institute of Theoretical Physics,
Academia Sinica} \centerline{\it P. O. Box 2735, Beijing 100080}
\medskip
\centerline{\it $^2$The interdisciplinary center of theoretical
Studies, Academia Sinica} \centerline{\it P. O. Box 2735, Beijing
100080}
\medskip

\centerline{\it  $^3$Graduate University of the Chinese Academy of
Sciences, Beijing 100080, P.R. China}
\medskip
 \centerline{\tt
huangqg@itp.ac.cn} \centerline{\tt mli@itp.ac.cn} \centerline{\tt
jhshe@itp.ac.cn}

\bigskip

WMAP three-year data favors a red power spectrum at the level of 2
standard deviations, which provides a stringent constraint on the
inflation models. In this note we use this data to constrain brane
inflation models and find that KKLMMT model can not fit WMAP+SDSS
data at the level of 1 standard deviation and a fine-tuning, eight
parts in thousand at least, is needed at the level of 2 standard
deviation.

\Date{April, 2006}

%\draft

\nref\infl{A.H. Guth, Phys.Rev.D 23(1981)347; A.D. Linde,
Phys.Lett.B 108(1982)389; A. Albrecht and P.J. Steinhardt,
Phys.Rev.Lett. 48(1982)1220. }

\nref\wmapts{D.N. Spergel et. al, astro-ph/0603449.}

\nref\al{L. Alabidi and D.H. Lyth, astro-ph/0603539. }

\nref\pe{H. Peiris., R. Easther, astro-ph/0603587; R. Easther, H.
Peiris, astro-ph/0604214. }

\nref\adl{A.D Linde, Phys.Lett.B 129(1983)127. }

\nref\dkmw{S. Dimopoulos, S. Kachru, J. McGreevy and J. Wacker,
hep-th/0507205;  R. Easther, L. McAllister, hep-th/0512102.}

\nref\dt{G.R. Dvali and S.H. Tye, Phys.Lett.B 450 (1999) 72,
hep-ph/9812483.}

\nref\brinf{S. Alexander, Phys.Rev.D 65 (2002) 023507,
hep-th/0105032; G. Dvali, Q. Shafi and S. Solganik, hep-th/0105203;
C. Burgess, M. Majumdar, D. Nolte, F. Quevedo, G. Rajesh and R.J.
Zhang, JHEP 07 (2001) 047, hep-th/0105204; G. Shiu and S.H. Tye,
Phys.Lett.B 516 (2001) 421, hep-th/0106274; D. Choudhury, D.
Ghoshal, D.P. Jatkar and S. Panda, hep-th/0305104. }

\nref\fq{F. Quevedo, hep-th/0210292. }

\nref\kklt{S. Kachru, R. Kallosh, A. Linde and S. Trivedi,
Phys.Rev.D 68(2003)046005, hep-th/0301240. }

\nref\kklmmt{S. Kachru, R. Kallosh, A. Linde, J. Maldacena, L.
McAllister and S. Trivedi, JCAP 0310(2003)013, hep-th/0308055. }

\nref\ft{H. Firouzjahi and S.H. Tye, hep-th/0501099. }

\nref\hk{Qing-Guo Huang and Ke Ke, Phys.Lett.B 633(2006)447,
hep-th/0504137. }

\nref\sh{S. Shandera, S.H. Tye, hep-th/0601099.}

\nref\bcsq{C.P. Burgess, J.M. Cline, H. Stoica and F. Quevedo, JHEP
0409(2004)033, hep-th/0403119.}

\nref\cs{ J.M. Cline, H. Stoica, Phys.Rev.D 72(2005)126004,
hep-th/0508029. }

\nref\bract{J.J. Blanco-Pillado, C.P. Burgess, J.M. Cline, C.
Escoda, M. Gomez-Reino, R. Kallosh, A. Linde, F. Quevedo,
hep-th/0603129. }

\nref\bhk{M. Berg, M. Haack, B. Kors, Phys.Rev. D71 (2005) 026005,
hep-th/0404087; M. Berg, M. Haack., B. Kors, JHEP 0511(2005)030,
hep-th/0508043; M. Berg, M. Haack., B. Kors, Phys.Rev.Lett.
96(2006)021601.}

\nref\becker{K. Becker, M. Becker, M. Haack, J. Louis, JHEP 0206
(2002) 060, hep-th/0204254.}

\nref\dterm{C.P. Burgess, R. Kallosh, F. Quevedo, JHEP 0310 (2003)
056, hep-th/0309187.}

\nref\mass{S. E. Shandera, JCAP 0504 (2005) 011, hep-th/0412077;
K. Bobkov, JHEP 0505 (2005) 010, hep-th/0412239; L. McAllister,
JCAP 0602 (2006) 010, hep-th/0502001.}

\nref\sel{U. Seljak, A. Slosar, Phys.Rev. D74 (2006) 063523,
astro-ph/0604143.}

Inflation dynamically resolves many puzzles concerning the hot big
bang cosmology, such as homogeneity, isotropy and flatness of the
universe \infl, and moreover it generates naturally superhorizon
fluctuations without appealing to fine-tuned initial setup. These
fluctuations become classical after crossing out the Hubble
horizon. During the deceleration phase after inflation they
re-enter the horizon, and seed the matter and radiation
fluctuations observed in the universe. The microwave anisotropy
encodes information from inflation. The $\Lambda$CDM model remains
an excellent fit to the three-year WMAP result as well as other
astronomical data \wmapts.

The results of WMAP three-year data are presented in \wmapts. For
$\Lambda$CDM model, WMAP three year data only shows that the index
of the power spectrum satisfies
\eqn\wmapn{n_s=0.951^{+0.015}_{-0.019};} combining WMAP with SDSS,
the result is\eqn\wmapf{n_s=0.948^{+0.015}_{-0.018}, } at a level of
1 standard deviation. A red power spectrum is favored at least at
the level of 2 standard deviations. If there is running of the
spectral index, the constraints on the spectral index and its
running and the tensor-scalar ratio are
\eqn\arn{n_s=1.21^{+0.13}_{-0.16}, \quad \alpha_s={dn_s \over d \ln
k}=-0.102^{+0.050}_{-0.043}, \quad r\leq 1.5.} What's important is
that WMAP three year data does not favor a scale-invariant spectrum
of fluctuations any more. While allowing for a running spectral
index slightly improves the fit to the WMAP data, the improvement in
the fit is not significant enough to require the running. WMAP group
claimed that the simple chaotic inflation model with potential
$m^2\phi^2$ \adl\ fits the observations very well. The WMAP three
year data provides significant constraints on the inflation models
and has ruled some inflation models out in \al. Other discussions
are given in \pe.

In spite of many phenomenological successes of inflation, there
nevertheless exist many serious problems, such as the singularity
problem, trans-Plankian problem, as well as questions concerning the
origin of the scalar field driving inflation. These problems can
only be addressed in a framework broader than effective field
theory. And it is expected that quantum theory of gravity should be
employed, for which string theory is the only self-consistent
scenario till now. So it is important to try to embed the many
inflation models into string theory. However this is not easy. Even
the above mentioned simple $m^2\phi^2$ inflation model has not yet
been realized in string theory. \foot{The authors in \dkmw\ proposed
an interesting idea to realize an assisted chaotic-like inflation,
called N-flation, by using a large number of axion fields in string
theory. } In the recent few years, much advance has been achieved in
this direction. One possible inflation scenario which has a natural
set up in string theory is driven by the potential between the
parallel dynamical brane and anti-brane, namely brane inflation
\refs{\dt-\fq}. However it is generally not easy to get a
sufficiently flat inflaton potential in brane inflation models \fq.
Kachru et al. in \kklt\ successfully introduced some
$\overline{D}3$-branes in a warped geometry in type IIB superstring
theory to break supersymmetry and uplift the AdS vacuum to a
metastable de Sitter vacuum with lifetime long enough. Putting an
extra pair of brane and anti-brane in this scenario, a more
realistic slow roll inflation, named KKLMMT model, is naturally
realized \kklmmt. KKLMMT inflation model has also been discussed in
\refs{\ft,\hk,\sh} in detail, where a possible conformal-like
coupling between the scalar curvature and the inflaton is taken into
account and a fine tuning at the level of roughly 1 part in ten is
needed. See \bcsq\ \cs\ for more considerations about KKLMMT model.

In this short note, we investigate the constraints on the brane
inflation model by using the WMAP three year data. We find that
KKLMMT model is not good at fitting the new data. It seems that
the better racetrack inflation is more likely to serve such role
\bract.

First we consider some general brane inflation models. We start with
a pair of $Dp$ and $\overline{D}p$-branes $(p\geq 3)$ filling the
four large dimensions and seperated from each other in the extra six
dimensions which are compactified. The D-brane tension provides an
effective cosmological constant $V_0$ on the brane, inducing an
inflation. The separation of the branes serves as the inflaton with
potential \eqn\pbf{V=V_0\(1-{\mu^n \over \phi^n} \).} The second
term in \pbf\ comes from the attractive interaction between the
branes and $n=d_\perp-2$, where $d_\perp=9-p$ is the number of the
transverse dimensions. Since $d_\perp \leq 6$, $n\leq 4$. At the
moment with the number of e-folds $N$, the inflaton field $\phi$ has
the value \eqn\pbpn{\phi_N=\(n(n+2)NM_p^2\mu^n \)^{1\over n+2}, }
where $M_p$ is the reduced Planck mass in four dimensions. The
corresponding slow-roll parameters are
\eqn\bfsrp{\eqalign{\epsilon_v&\equiv{M_p^2\over 2}\({V'\over V}
\)^2={n^2 \over 2\(n(n+2) \)^{2(n+1)\over n+2}}N^{-{2(n+1)\over
n+2}}{\mu^{2n\over n+2}\over M_p^{2n\over n+2}}, \cr
\eta_v&\equiv{M_p^2{V'' \over V}}=-{n+1\over n+2}{1\over N}, \cr
\xi_v&\equiv M_p^4{V'V'''\over V^2}={n+1\over n+2}{1 \over N^2}. }}
In general, both $V_0$ and $\mu^4$ are roughly the effective brane
tension compactified to four dimensions, and thus much smaller than
$M_p^4$. So the amplitude of the tensor perturbations is negligible
\fq. The spectral index and its running become
\eqn\pbsr{\eqalign{n_s&=1-{n+1\over n+2}{2\over N},\cr
\alpha_s&=-{n+1\over n+2}{2\over N^2}.}} For $N\sim 50$, the running
of spectral is roughly $-{\cal O}(10^{-3})$ which is also
negligible. The spectral index for different power $n$ and number of
e-folds $N$ are showed in fig. 1.

\bigskip
{\vbox{{
        \nobreak
    \centerline{\epsfxsize=7cm \epsfbox{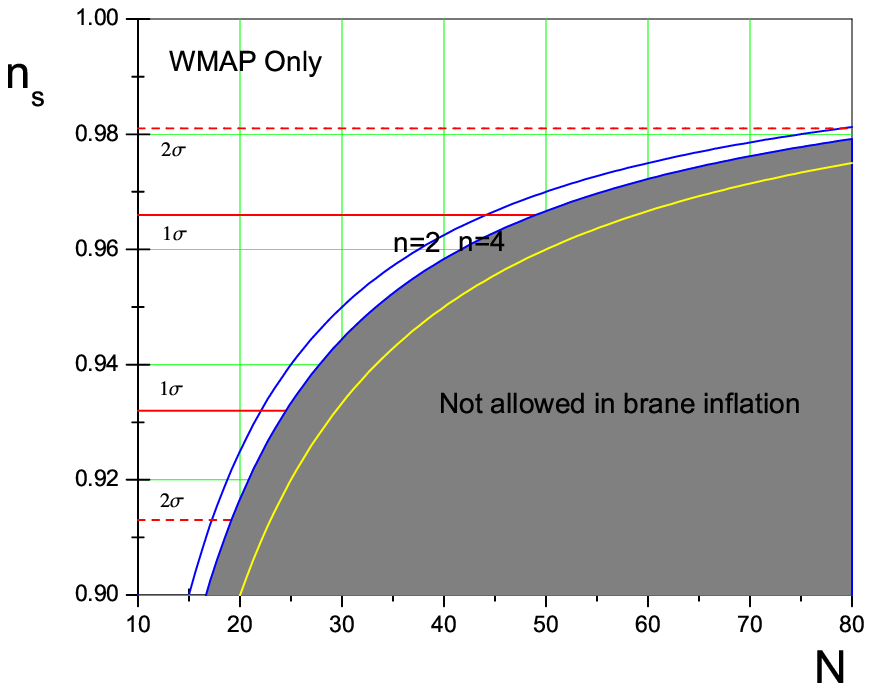}
    \epsfxsize=7cm \epsfbox{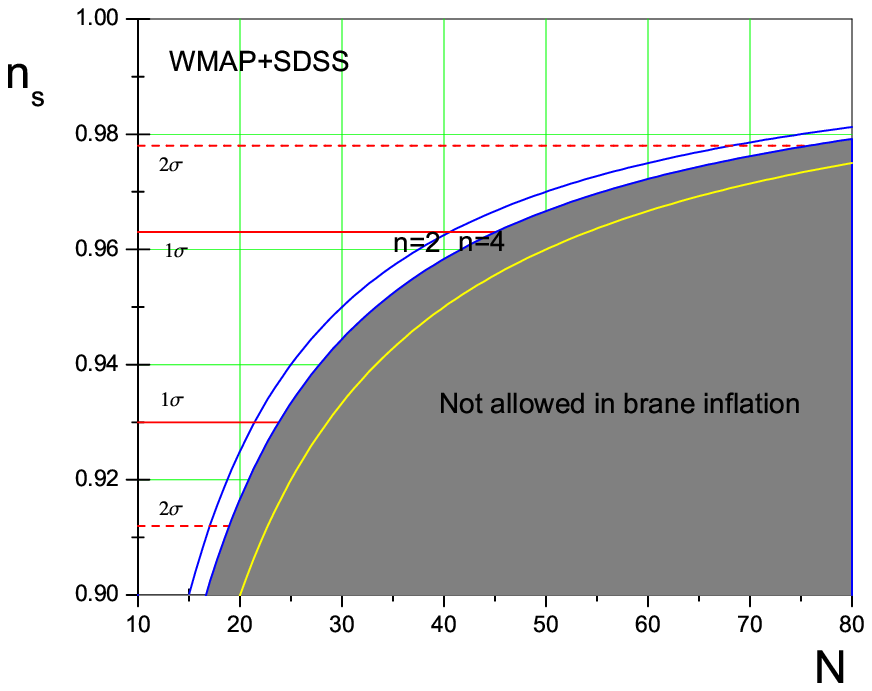}}
        \nobreak\bigskip
    {\raggedright\it \vbox{
{\bf Figure 1.} {\it The yellow lines correspond to $n\rightarrow
\infty$. The region with $n>4$ is not allowed, since the number of
the brane transverse dimension is not greater than 6. }
 }}}}
    \bigskip}

We see from the above figure that the smaller the number of the
transverse dimensions, the worse the fitting with WMAP data. For
$n=2,4$ (corresponding respectively to the D5 and D3 brane cases),
the fitting results for the number of e-folds for WMAP only are
\eqn\gnewo{\eqalign{N=30.6^{+13.5+48.3}_{-8.50-13.4} \quad
\hbox{for} \quad n=2, \cr N=34.0^{+15.0+53.7}_{-9.50-14.9} \quad
\hbox{for} \quad n=4; }} for WMAP+SDSS
\eqn\gnews{\eqalign{N=28.8^{+11.7+39.4}_{-7.4-11.8} \quad \hbox{for}
\quad n=2, \cr N=32.0^{+13.0+43.8}_{-8.20-13.1} \quad \hbox{for}
\quad n=4. }} On the other hand, the authors in \al\ proposed the
relevant e-folding number corresponding to the observation is
\eqn\refd{N=54\pm 7.} The brane inflation model for $n=4$ is near
the boundary of the WMAP three years data only and can not fit the
data of WMAP+SDSS at the level of $1\sigma$; the case with $n=2$ is
outside the allowed range by WMAP only or WMAP+SDSS at the level of
$1\sigma$. But they still survive at the level of $2\sigma$.

We neglect a problem for the brane inflation model in the above
discussions, which says that the distance between the brane and
anti-brane must be larger than the size of the extra dimensional
space if the inflation is slow rolling, or $\eta$ sufficiently
small, in this scenario. Maybe the potential in eq. \pbf\ can emerge
in other theories we have not known.

A more realistic embedding of brane inflation into string theory is
the so called KKLMMT model \kklmmt. They consider in type IIB string
theory highly warped compactifications with all moduli stabilized by
the combination of fluxes and nonperturbative effects. With a small
number of $\overline{D}3$-branes added, the vacuum is lifted to dS.
They consider further in the warped throat a D3-brane moving towards
the $\overline{D}3$-brane located at the bottom of the throat, thus
realizing the scenario of brane/antibrane inflation. The warping of
the geometry also provides a natural mechanism to achieve a
sufficiently flat potential and thus a sufficiently small $\eta$,
since the warped geometry gives rise to a redshift which reduces the
effective tension of anti-brane and the potential coming from the
attractive force between brane and anti-brane is redshifted to be
much smaller than the effective tension term \kklmmt. In this note,
we consider the potential for KKLMMT model which takes the form, in
\ft, \eqn\gkp{V=\half \beta H^2 \phi^2+2T_3h^4\(1-{\mu^4 \over
\phi^4} \), } where \eqn\td{T_3={1 \over (2\pi)^3g_s\ap^2}} is
D3-brane tension, $h$ is the warped factor in the throat and
\eqn\dm{\mu^4={27\over 32\pi^2}T_3h^4.} The newly added $\phi^2$
term comes from the K\"{a}hler potential, D-term and also
interactions in the superpotential, and $H$ is the Hubble constant.
Generally $\beta$ is of order unity \kklmmt, but to achieve slow
roll, it has to be fine-tuned to be much less than one \refs{\ft,
\hk}. In the following, we will proceed to further constrain $\beta$
using the WMAP three year data.

Inflation is dominated by the D3-brane tension. With the number of
e-folding equals $N$ before the end of inflation, the inflaton
field is \eqn\pn{\phi_{N}^6=24 N M_p^2 \mu^4 m(\beta), } where
\eqn\mb{m(\beta)={(1+2 \beta)e^{2 \beta N} -(1+\beta/3) \over 2
\beta (N+5/6) (1+\beta/3)}. } Now the slow roll parameter can be
expressed as \eqn\srp{\eqalign{\epsilon_v&={1 \over 18}
\left({\phi_{N} \over M_p} \right)^2 \left(\beta+{1 \over 2 N
m(\beta)} \right)^2, \cr \eta_v&={\beta \over 3} -{5 \over 6}{1
\over N m(\beta)}, \cr \xi_v&={5 \over 3Nm(\beta)}\(\beta+{1 \over
2Nm(\beta)} \). }} The amplitude of the the scalar comoving
curvature fluctuations has been calculated in detail
\refs{\ft,\hk}  \eqn\asd{\Delta_{\cal R}^2\simeq {V/M_p^4 \over
24\pi^2\epsilon_v}=\(2^5 \over 3\pi^4\)^{1/3} \({T_3h^4\over
M_p^4}\)^{2 \over 3} N^{5\over 3}f^{-{4 \over 3}}(\beta), } where
\eqn\fb{f(\beta) = m^{-{5/4}}(\beta)(1+2\beta N m(\beta))^{3\over
2}. } Substituting eq. \dm\ and \asd\ into \pn, we have
\eqn\pmp{{\phi_{N} \over M_p}=\left({27 \over 8} \right)^{1\over
4} m^{1 \over 6}(\beta) f^{1 \over 3} (\beta) N^{-{1\over 4}}
\(\Delta_{\cal R}^2\)^{1\over 4}. } Now we have
\eqn\epp{\epsilon_v={1 \over 4\sqrt{6N}}\(\Delta_{\cal
R}^2\)^\half m^{1\over 3}(\beta)f^{2\over 3}(\beta)\(\beta+{1\over
2Nm(\beta)}\)^2.} Usually the normalization of the primordial
scalar power spectrum is $\Delta_{\cal R}^2 \simeq 2\times
10^{-9}$ for $N\sim 50$. The spectral index and its running are
\eqn\ind{\eqalign{n_s&=1-6\epsilon_v+2\eta_v, \cr
\alpha_s&=-24\epsilon_v^2+16\epsilon_v\eta_v-2\xi_v,}} and the
tensor-scalar ratio is \eqn\tsr{r=16\epsilon_v.} According to eq.
\srp, the $\beta$ term in the potential makes the power spectrum
trending to be blue.

For $\beta\leq 0.1$, the tensor-scalar ratio is roughly $10^{-9}
\sim 10^{-5}$ and thus the tensor perturbations are negligible.
Now the spectral index and its running are
\eqn\sbdr{\eqalign{n_s&=1+{2\beta \over 3}-{5 \over 3m(\beta)}{1
\over N}, \cr \alpha_s&=-{10\beta \over 3m(\beta)}{1 \over N}-{5
\over 3m^2(\beta)}{1\over N^2}.}} The spectral index is shown in
fig. 2. The running of the spectral index is roughly $-{\cal
O}(10^{-3})$ and is negligible.

\bigskip
{\vbox{{
        \nobreak
    \centerline{\epsfxsize=7cm \epsfbox{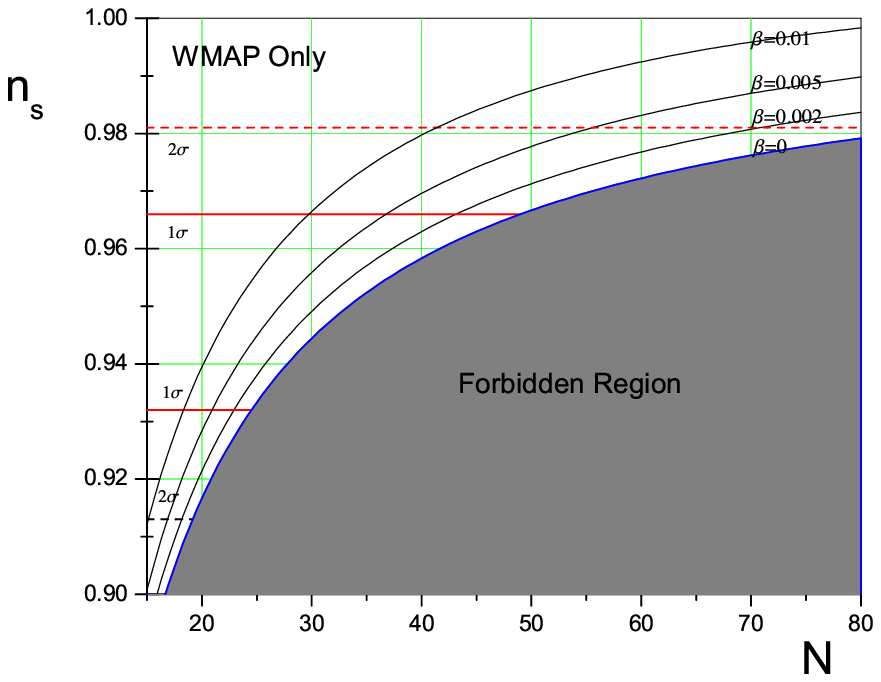}
    \epsfxsize=7cm \epsfbox{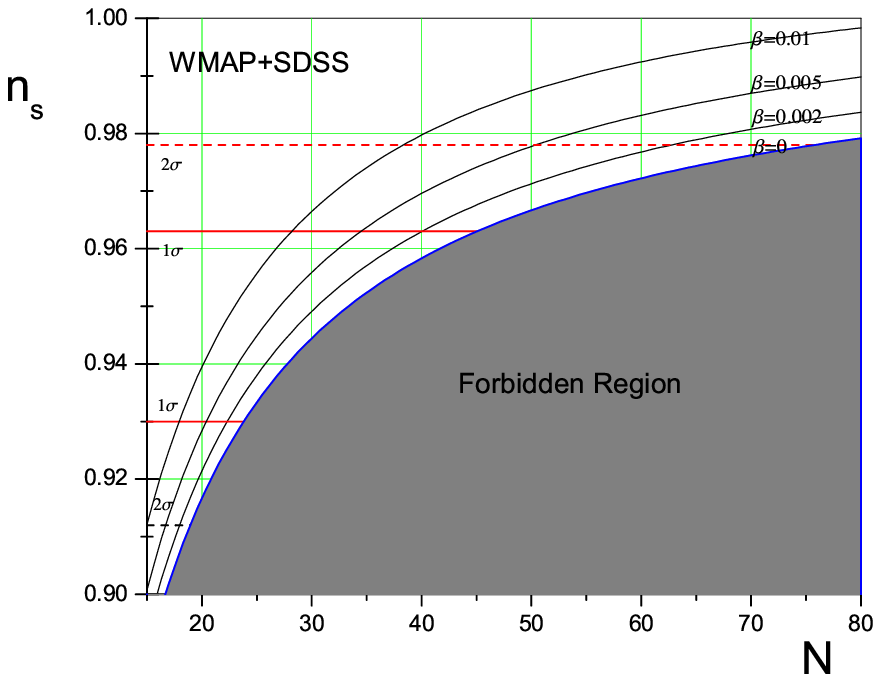}}
        \nobreak\bigskip
    {\raggedright\it \vbox{
{\bf Figure 2.} {\it The tensor perturbations can be neglected. If
$\beta<0$, the $\beta$ term in the potential trends to push D3-brane
out of the throat and the inflation does not happen. }
 }}}}
    \bigskip}

For $\beta=0$, the fitting result is just the case for $n=4$ in
eq. \gnewo\ and \gnews. The KKLMMT model is on the boundary at the
level of $1\sigma$ of the WMAP three years data only and does not
fit the data of WMAP+SDSS. Larger the value of $\beta$, larger the
spectral index. Thus the $\beta$ term in the potential can not
improve the fitting results. The constraints on the parameter
$\beta$ are $\beta \leq 6\times 10^{-4}$ at the level of $1\sigma$
and $\beta \leq 8\times 10^{-3}$ at the level of $2\sigma$ for
WMAP only; $\beta \leq 6\times 10^{-3}$ at the level of $2\sigma$
for WMAP+SDSS. The fine tuning for the parameter $\beta$ is
needed!

For $\beta>0.1$, a large amplitude of the tensor perturbations
emerges and a large running is possible. If the running of the
spectral index is large enough, a blue power spectrum is allowed
\arn. We show the spectral index and its running and the
tensor-scalar ratio for different $\beta$ in fig. 3.

\bigskip
{\vbox{{
        \nobreak
    \centerline{\epsfxsize=5cm \epsfbox{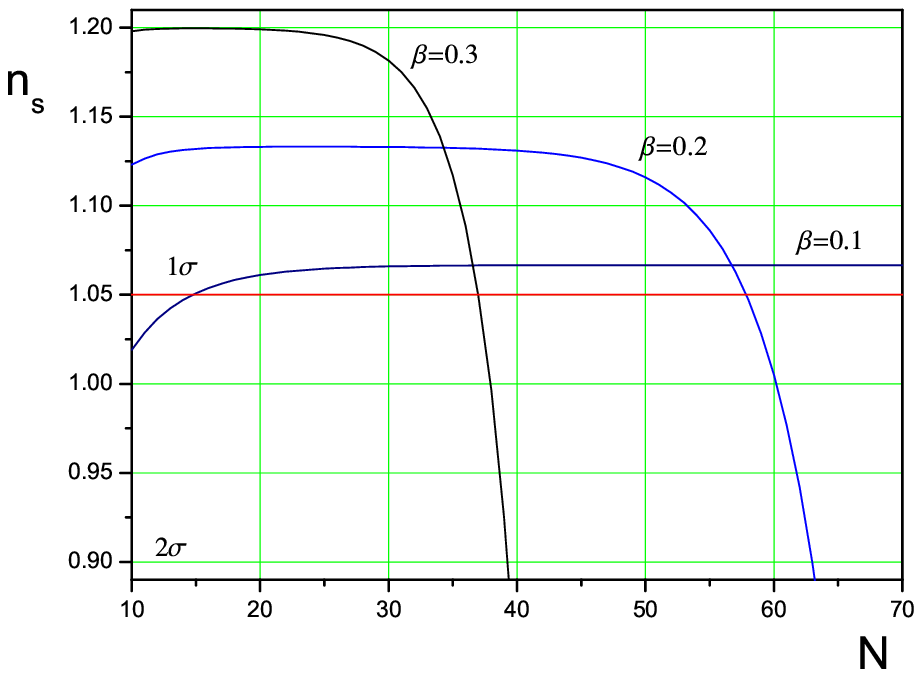}
    \epsfxsize=5cm \epsfbox{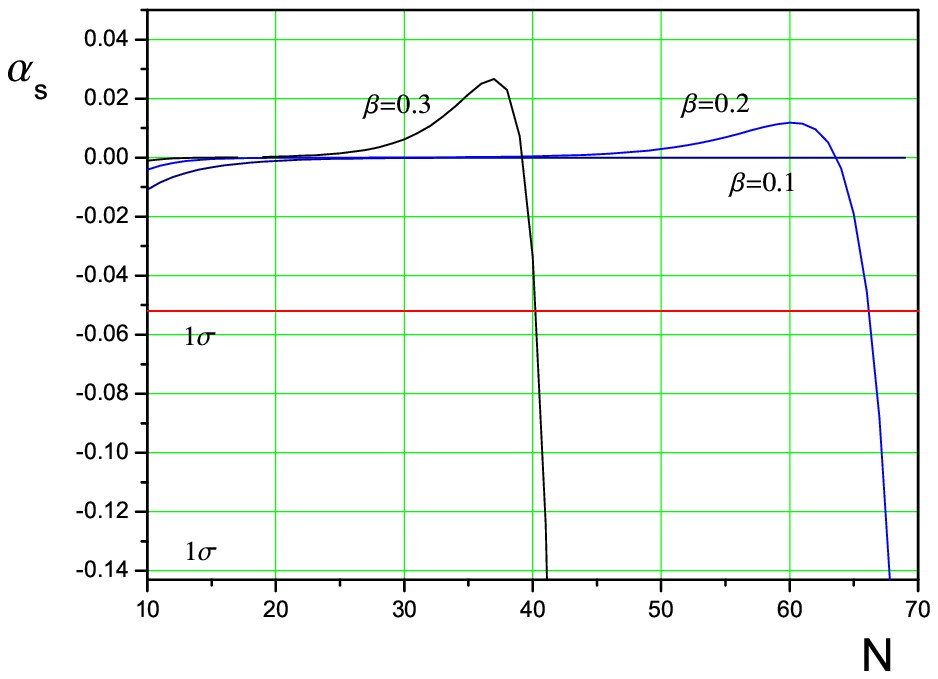}
    \epsfxsize=5cm \epsfbox{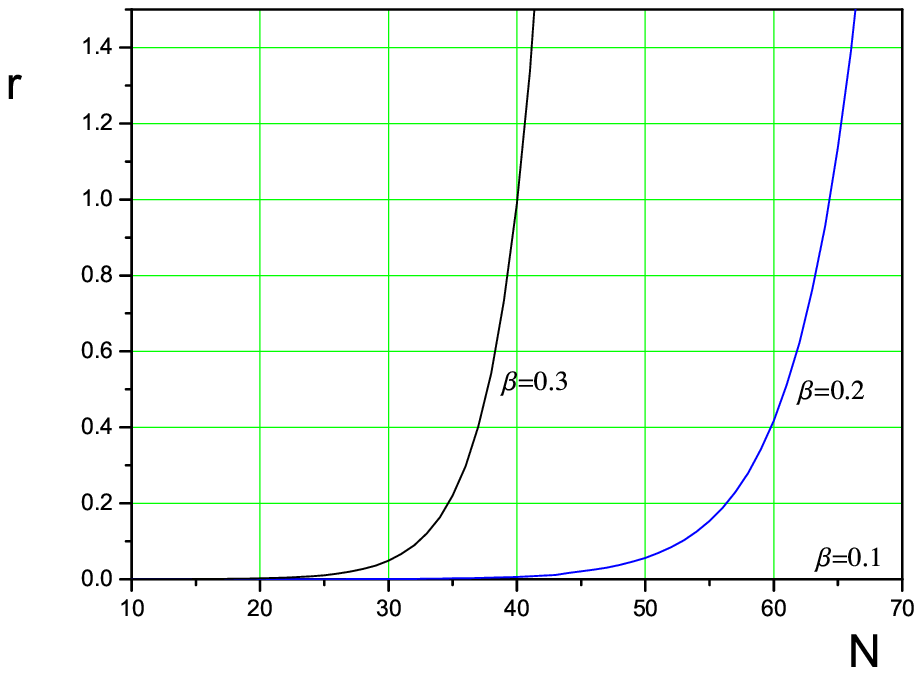}}
    \nobreak
    \centerline{\epsfxsize=5cm \epsfbox{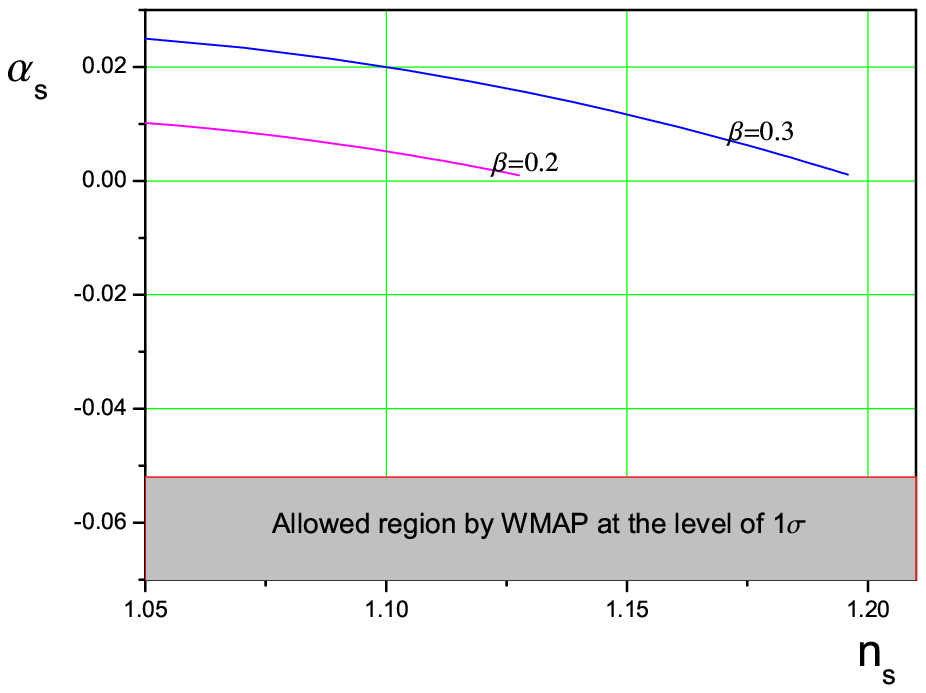}}
                 \nobreak\bigskip
    {\raggedright\it \vbox{
{\bf Figure 3.} {\it The top three figures are the spectral index
and its running and the tensor-scalar ratio respectively vary with
the number of e-folds for different $\beta$. The bottom figure shows
the correlation between the spectral index and its running.}
 }}}}
    \bigskip}

Unfortunately, the bottom figure in fig 3 shows that the KKLMMT
model can not provide a reasonable running of the spectral index.

To summarize, WMAP three year data provides a stringent constraint
on the brane inflation. KKLMMT model can not fit the new data
well. In order to stabilize the moduli as in \kklt, the mass of
the inflaton is roughly the same as the Hubble parameter during
the period of inflation and thus the inflation is impossible. We
need to fine-tune the parameter $\beta$ to reduce the mass of the
inflaton. Our fitting results show that a fine-tuning, eight parts
in thousand at least, is needed at the level of $2\sigma$. It
seems that the better racetrack inflation proposed in \bract\ is
favored by WMAP three-year data.

One may go on to ask whether such stringent fine-tuning is
possible in KKLMMT-type models. Thus we need to look further into
the origin of $\beta$. In KKLMMT's original model, with
superpotential \eqn\super{W=W_0+g(\rho)f(\rho),} where $W_0$ comes
from the fluxes, $f(\rho)=1+\delta\phi^2$, $\delta$ a constant,
and $g(\rho)$ an arbitrary function of $\rho$, the parameter
$\beta$ has the form \kklmmt \eqn\bbeta{\beta=1-{{\mid
V_{AdS}\mid}\over{V_{dS}}}(\gamma-2\gamma^2),} with
$\gamma=\delta{g\over{g^{'}}}.$ Lack of precise knowledge of the
superpotential $W(\rho,\phi)$, when \eqn\vds{\mid V_{AdS}\mid \gg
V_{dS},} fine-tuning can be performed to get any value for
$\beta$. For $T^2\times T^4/Z_2$ and $T^6/Z_N$ models, the $\ap$
collection to the superpotential is calculated \bhk, and
correction to the inflaton mass is thus determined with the result
\eqn\cbeta{\beta=1-{{\mid V_{AdS}\mid}\over{V_{dS}}}\Delta,} where
$\Delta\simeq 0.1$. There is still enough room for fine-tuning.
Moreover, inflaton mass also receives contributions from the
K\"{a}hler potential \becker\ and D-terms \dterm. For further
discussion of inflaton mass problem see \mass\ and references
therein.

To solve the fine tuning problem in the KKLMMT scenario, Cline and
Stoica proposed a novel dynamical mechanism in \cs. They introduced
an additional parameter $\psi_0$, which is approximated as zero in
KKLMMT's original paper, the position of the antibrane relative to
the equilibrium position of the brane in the absence of the
antibrane. They further considered the presence of multiple mobile
branes. With sufficiently large number of branes, the inflaton
potential has a metastable minimum, where the branes are initially
confined. These branes can tunnel out of the minimum, making the
minimum shallower and shallower, and the potential more and more
flat. Finally the remaining branes will roll together into the
throat, driving the inflation. In this way, a sufficiently flat
inflaton potential is achieved through a dynamical mechanism,
without the need of fine-tuning.

In their scenario, the inflaton potential is modified to be
\eqn\csip{V={{TN}\over
A^3}(1-{1\over6}\phi^2)^{-2}(1-{{3b(N/A)^3}\over{(\phi-\phi_0)^4}})^{-1},}
with $\phi$ the canonically normalized inflaton field, $\phi_0$ the
canonically normalized value of $\psi_0$ and N the number of branes.
While $T=2\tau/g_s^4$, with 3-brane tension $\tau$ and string
coupling $g_s$. The properly combined parameter
$A=(2\sigma)^{2/3}\epsilon^{-1/3}$, with warp factor $\epsilon$ and
K\"{a}hler modulus $\sigma$.

The scalar spectral index at 55 e-foldings before the end of
inflation is calculated in \cs\ to be $0.93<n_s<1.15$, while
correspondingly the running of the spectral index is $-0.012<dn/d\ln
k<-0.001$. In the range permitted by WMAP 3-year data without
running, where $0.93<n_s<0.963$, their prediction of running is
$-0.003<dn/d\ln k<-0.001$, small enough to be negligible.

In conclusion, Cline and Stoica's scenario \cs\ removes the
fine-tuning problem of the brane inflation models, and fits well
with the WMAP 3-year result without running. But their scenario can
not produce a large enough running to fit the WMAP 3-year result
with running.

Recently, there is also discussion of KKLMMT scenario in light of
WMAP 3-year data in \sel, where they seek string effects from CMB B
polarization. They expect their predictions to be general and do not
depend on the exact form of the inflaton potential. So they do not
consider the constraints of $\beta$ as we do above, and simply
choose a fixed value $\beta=0.02$.

\bigskip

Acknowledgments.

We thank Y. Li and Z. G. Xiao for helping with the drawings, and D.
W. Pang for carefully reading the draft. We also thank J. Cline and
H. Stoica for correspondence on their paper. The work of QGH was
supported by a grant from NSFC, a grant from China Postdoctoral
Science Foundation and and a grant from K. C. Wang Postdoctoral
Foundation. The work of ML and JHS was supported by a grant from CAS
and a grant from NSFC.

\listrefs
\end